\documentclass[preprint]{aastex}
\usepackage{float}

\newcommand{\sgr}{\mbox{SGR\,J0501$+$4516~}}
\newcommand{\sgrnos}{\mbox{SGR\,J0501$+$4516}}

\shorttitle{Link of the persistent and burst X-ray emission of \sgrnos}
\shortauthors{Lin et al.}

\begin{document}

\title{On the X-ray emission mechanisms of the persistent source and very low-fluence bursts of \sgrnos}

\author{Lin Lin\altaffilmark{1}, 
	Ersin G\"o\u{g}\"u\c{s}\altaffilmark{1},
	Tolga G\"uver\altaffilmark{1},
	Chryssa Kouveliotou\altaffilmark{2}
	}

\email{linlin@sabanciuniv.edu}

\altaffiltext{1}{Sabanc\i~University, Faculty of Engineering and Natural Sciences, Orhanl\i$-$ Tuzla, \.{I}stanbul 34956,
Turkey}
\altaffiltext{2}{Science and Technology Office, ZP12, NASA/Marshall Space Flight Center, Huntsville, AL 35812, USA}

\begin{abstract}

We present here a detailed spectral study of the X-ray emission of the persistent source and the low-fluence bursts of \sgr observed during a deep \textit{XMM-Newton} observation near the peak of its 2008 outburst. For the persistent emission we employ a physically motivated continuum emission model and spectroscopically determine important source properties; such as, the surface magnetic field strength and the magnetospheric scattering optical depth. We find that the magnetar surface temperature near the peak of its activity is $0.38$ keV, corresponding to an emission area of 131\,km$^2$ at a distance of 2\,kpc. The surface magnetic field strength determined spectroscopically, $B=2.2\times10^{14}$ G,  is consistent with the dipole field strength inferred from the source spin and spin down rate. We fit the stacked spectra of 129 very faint bursts with a modified blackbody model and find a temperature of $1.16$ keV, corresponding to an emission area of 93\,km$^2$. We also find an evidence for cooling during the burst decay phase. 

\end{abstract}


\section{Introduction}

Magnetars are manifestations of isolated neutron stars. Commonly known as Soft Gamma Repeaters (SGRs) and Anomalous X-ray Pulsars (AXPs), magnetars are generally characterized by relatively slow spin periods (in a narrow range of $2-12$\,s), rapid spin-down rates ($5\times10^{-13}-10^{-10}$\,s\,s$^{-1}$), and intense X-ray bursts with luminosities anywhere between $10^{37}-10^{45}$\,erg\,s$^{-1}$. Dipole magnetic field strengths of these sources, as inferred from their spin periods and spin-down rates, are in the order of $\sim10^{14}$\,G \citep{ck1998,ck1999}, which are in perfect agreement with the predictions of the magnetar model that there exist isolated neutron stars, which are powered by their own extremely strong magnetic fields \citep{dt1992,td1993}. 

Magnetars are persistent X-ray emitters in quiescence most of the time at luminosity levels around $\sim10^{33}-10^{36}$\,erg\,s$^{-1}$. As they enter active episodes, they usually exhibit significant variations in their persistent emission spectral and temporal behavior in conjunction with the outburst onset \citep{rea2011}. Transient magnetars are generally undetectable unless they undergo outbursts, during which their X-ray fluxes are enhanced by factors of a few $10-100$, followed by a gradual decay over timescales of months to years \citep[e.g.][]{gogus2010,rea2011}. Typical magnetar bursts are short, lasting only for $\sim100$\,ms, with a peak luminosity lower than $\sim10^{41}$\,erg\,s$^{-1}$ \citep{gogus2001,vdh2012}. More energetic bursts occur much less frequently and have been observed only from five SGR sources \citep{mazets1979,hurley1999,hurley2005,mazets1999,watts2010,gogus2011}. For reviews on magnetars, see \citet{woods2006} and \citet{mereghetti2008}.

Both persistent X-ray emission and bursts are attributed to the decay of extremely strong magnetic fields in the framework of the magnetar model \citep{td1993,td1995,thompson2002}. However, it is not trivial to describe the observed data with detailed physical models, considering that the radiation ought to emerge from the strong magnetic and gravitational fields. The persistent X-ray spectrum of magnetars is usually fitted with phenomenological models: either the combination of a blackbody and a power law \citep[see section 14.3.1 of][]{woods2006}, or two blackbody components \citep{mereghetti2008}. Recently, \citet{guver2007,guver2008} and \citet{gogus2011} studied the persistent emission spectra with an idealized physical model, the Surface Thermal Emission and Magnetospheric Scattering (STEMS). The STEMS model accounts for the transport of radiation in a fully ionized and extremely magnetized atmosphere \citep{ozel2001,ozel2003} as well as for resonant cyclotron scattering (RCS) \citep{lyutikov2006} of photons emerging from the surface in the magnetosphere of the magnetar (see also Section \ref{sec:mo}). A typical magnetar burst mainly emits radiation up to 100\,keV. Its spectrum is well described phenomenologically by the sum of two blackbody functions with temperatures of $\sim$3 and $\sim$12 keV \citep[e.g.][]{israel2008,lin2012}. To physically describe the burst spectra, \citet{lyubarsky2002} introduced analytically the shape of a thermalised spectrum modified by a strong magnetic field ($\sim10^{14}$\,G). We use here the latter model in our burst spectral analysis. 

\sgr emitted a series of short X-ray/soft $\gamma-$ray bursts for almost two weeks starting on 2008 August 22; several tens of events were detected with \textit{Swift}, \textit{Fermi}/GBM, \textit{Konus-Wind}, and \textit{Suzaku} \citep{enoto2009,aptekar2009,kumar2010,nakagawa2011,lin2011}. A pointing observation with \textit{RXTE} revealed a spin period of $5.762$\,s. Using the long term monitoring with \textit{RXTE} and \textit{Swift}/XRT, a spin-down rate of $\sim5.8\times10^{-12}$\,s\,s$^{-1}$ was later obtained providing an estimate of the inferred dipole magnetic field strength of $\sim2\times10^{14}$\,G \citep{gogus2010}. \textit{Chandra} observations of \sgr located the point source very precisely at $R.A. = 05^{\rm h} 01^{\rm m} 06\fs76$, $Dec = +45\arcdeg 16\arcmin 33\farcs92$ (J2000), with a 1$\sigma$ uncertainty of $0\farcs11$ \citep{gogus2010}. Unlike many other galactic magnetars, \sgr is located in the anti-Galactic center direction, most likely at the Perseus arm at $\sim2$\,kpc \citep{xu2006}. As a result, the interstellar Hydrogen column density, $N_{\rm H}$, towards \sgr is much less compared to the values of magnetars located in the Galactic center direction, making \sgr an ideal source for studying its soft X-ray emission characteristics. 

In this paper we introduce for the first time the RCS process into the modified thermal burst emission, to study the burst spectra below 10\,keV from \sgrnos. We also analyze the persistent emission of \sgr with the STEMS model, and investigate whether there exists a link between the persistent emission and the dim burst spectra assuming the same scattering process. We describe the physical models used for the persistent emission and burst spectra in Section \ref{sec:mo}. In Section \ref{sec:obs}, we describe the analysis of the \textit{XMM-Newton} data used here, and present the results of our spectral studies in Section \ref{sec:specresult}. Finally, we discuss the interpretation of these results in Section \ref{sec:disc}.

\section{Models for the Persistent and Burst Emission\label{sec:mo}}

\subsection{Persistent emission}

In the magnetar model, the quiescent X-ray emission could originate from the neutron star surface heating by the decay of the strong magnetic field \citep{td1993}. Persistent emission photons ought to go through the atmosphere and magnetosphere of the highly magnetized neutron star before they escape. The transport of radiation for different polarization modes was investigated in detail in a series of studies considering the effects of absorption, emission, and scattering by the fully ionized plasma \citep{zane2001,ozel2001,ho2001,ho2003,ozel2003}. These showed that as photons propagate from the inner to the outer layers of a magnetized atmosphere, the spectrum of the emergent emission becomes harder than the original blackbody shape. 

As the photons reach the magnetosphere, resonant cyclotron scattering could effectively modify their spectrum  \citep{thompson2002}. This process will again make the spectrum harder, by scattering the low energy photons to higher energies. \citet{lyutikov2006} calculated the simplified one dimensional resonant cyclotron scattering of a Planck spectrum by a non-relativistic warm plasma in an inhomogeneous magnetic field. Applying the same scattering process to the emission from the atmosphere, \citet{guver2007,guver2008} developed the Surface Thermal Emission and Magnetospheric Scattering model (STEMS). They then used this model to successfully describe the persistent emission spectra of XTE\,J1810$-$197 and 4U\,0142$+$61, and also estimate the surface magnetic fields of these two magnetars. The STEMS model depends on four parameters: the surface magnetic field ($B$), the temperature of the neutron star ($kT$), the temperature of the plasma in the magnetosphere ($\beta$), and the optical depth for resonant cyclotron scattering ($\tau$). Because of the strong gravitational field of the neutron star, the STEMS model also includes a fixed gravitational redshift ($z$) parameter, which depends only on the ratio of mass over radius of a neutron star. We fixed $z=0.306$ throughout this paper, corresponding to a typical neutron star with mass of $1.4\,M_\odot$ and radius of 10\,km.

\subsection{Burst emission}

One of the mechanisms invoked for the origin of the short duration magnetar bursts is the sudden release of energy by cracking of the magnetically strained neutron star crust \citep{td1995}. This sudden energy release creates a hot pair plasma, which then becomes trapped in the magnetosphere forming a bubble with large optical depth \citep{td1995}. The photons need to go through multiple scatterings to escape the bubble, thereby thermalizing the burst radiation \citep{td1995}. Unlike in the case of the persistent surface emission for which the absorption, emission and scattering are important, scattering is the most dominant process in the burst bubble \citep{lyubarsky2002}. Taking into account the fact that photons with different energies would have different scattering cross-sections, \citet{lyubarsky2002} calculated the radiation transfer for the magnetar bursts. The resulting spectral shape has a much flatter low energy distribution compared with the equilibrium blackbody spectrum (hereafter we call the resulting shape as the modified blackbody). The modified blackbody photons are also subject to multiple scatterings as they go through the magnetosphere. We adopted the same process of the resonant cyclotron scattering as used in \citet{lyutikov2006} and \citet{guver2007} to up-scatter the modified blackbody burst spectrum. The scattered modified blackbody and the Planckian function differ significantly at energies below the bolometric temperature ($T_b$). The softer emission is highly affected by the interstellar absorption, which would make the two models indistinguishable if the source were towards a high interstellar $N_{\rm H}$ region. The location of \sgr makes it, therefore, an excellent source to determine the nature of its burst emission mechanism. We generated a numerical grid for the modified blackbody with the resonant cyclotron scattering (MBB$+$RCS) to fit the burst spectral data in XSPEC \citep{arnaud1996}. The MBB$+$RCS model has three free parameters: the bolometric temperature of the bubble ($kT_b$) which was allowed to range between $0.1-20$\,keV, and $\beta$ and $\tau$ as defined in Section 2.1. We also used the same gravitational redshift parameter ($z=0.306$) while fitting the burst spectra. 

\section{Observations and Data Reduction \label{sec:obs}}

\sgr was observed with the \textit{XMM-Newton} Observatory \citep{jansen2001} in seven occasions between August and September 2008. Here, we selected only the first observation performed on 2008 August 23 (Observation ID: 0560191501), the most burst active day of the source, to study the spectral properties of the bursts and of the underlying persistent emission at the same time. The observation lasting for 48.9\,ks was performed in the small window mode of the European Photon Imaging Camera (EPIC) pn instrument \citep{struder2001}, allowing a temporal resolution of 6 ms. We processed the data using SAS version 11.0.0 with the latest calibration files generated on 2012 May 18.

We identified and removed events in the piled-up time intervals (64 s in total), which are the ones with count rates over 50 counts\,s$^{-1}$, from the source event list. We then constructed the source lightcurve with 100\,ms binsize as shown in Figure 1. The average count rate of the persistent emission is $\sim5.5$ counts\,s$^{-1}$, corresponding to  no more than 5.5 counts per bin in the light curve of Figure 1 (or a count rate of 55 counts\,s$^{-1}$). We assumed that a bin contained a burst if it exceeded the persistent X-ray emission level by at least $2\sigma$, corresponding to $\sim10$ counts per bin (or a count rate of 100 counts\,s$^{-1}$). We, therefore, accumulated the persistent emission spectrum using 100\,ms time bins with count rate less than 50 counts\,s$^{-1}$, and the burst spectra from those bins with over 100 counts\,s$^{-1}$. The total exposure time of the persistent emission spectrum is 32.7 ks. There are 129 identified bursts having 100\,ms peak count rates between 100 and 500 counts\,s$^{-1}$; the total time encompassed by these bursts was 8.7 s. We combined all 129 burst data into one stacked spectrum and regrouped the resulting spectra to have a minimum of 15 counts in each spectral bin, making sure that we kept the bin size within the spectral resolution of the instrument. We then fit the spectra using XSPEC version 12.7.0.


\begin{figure}[h]
    \includegraphics{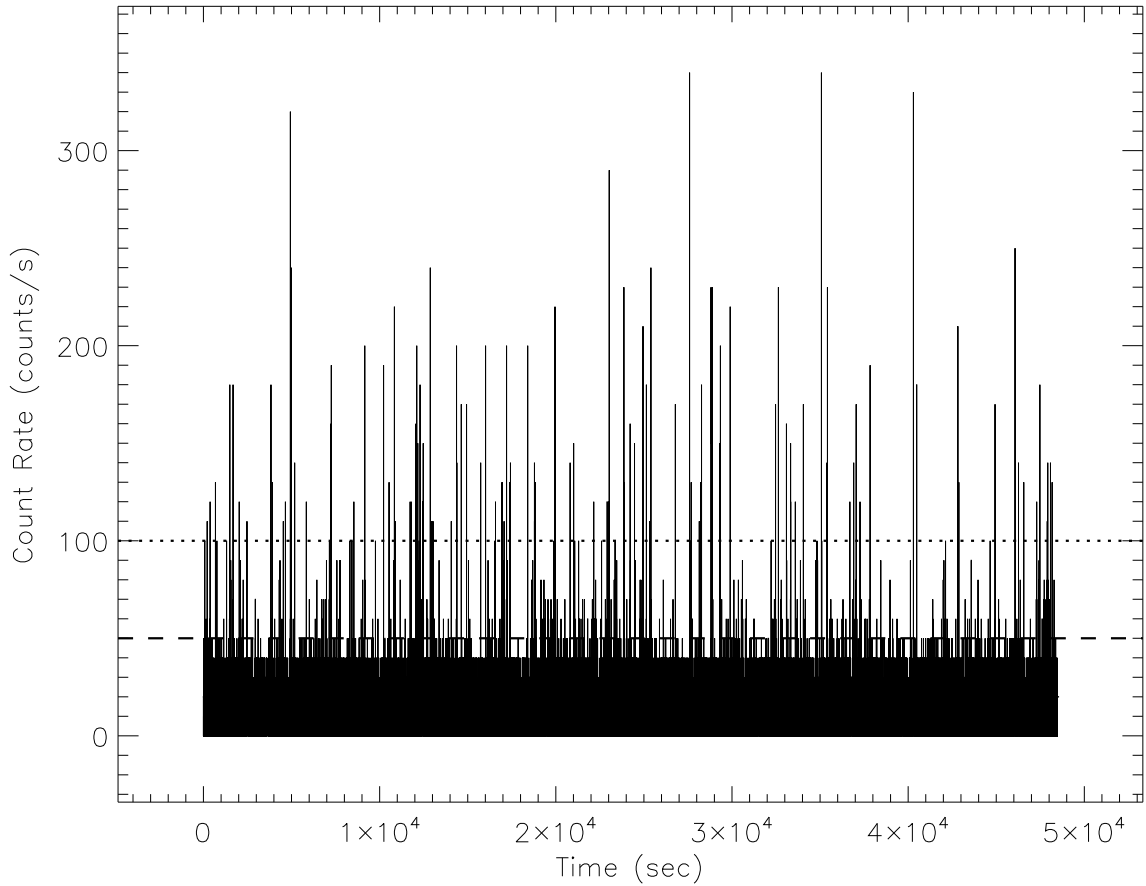}
\caption{The lightcurve of \sgr in 100 ms temporal resolution. The dashed line marks the count rate (50 counts\,s$^{-1}$) selected for the persistent emission level. The dotted line denotes the count rate threshold for the burst selection.  \label{100mslc}}
\end{figure}

\section{Spectral Analysis Results \label{sec:specresult}}

\subsection{The Persistent Emission Spectrum}

\begin{figure}[h]
    \includegraphics[angle=270, scale=0.55]{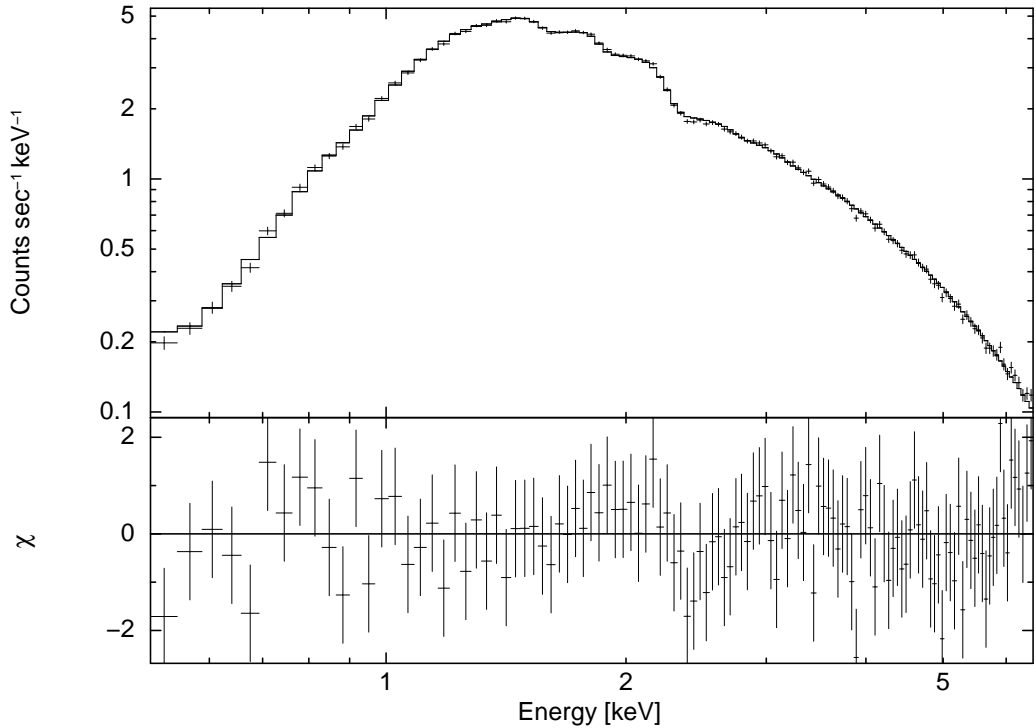}
\caption{The count spectrum of the persistent emission of \sgr overplotted with the best fit STEMS model curve (upper panel), and the fit residuals (lower panel).  \label{persis_spec}}
\end{figure}

We first fit the spectrum of the persistent emission with a single blackbody and a power law (BB$+$PL). The best fit parameters listed in Table \ref{specpar}, are consistent with the results reported in \citet{rea2009} and \citet{gogus2010}. We then fit the spectrum with STEMS (see also Figure \ref{persis_spec}). The STEMS model can fit the persistent spectrum equally well as the BB$+$PL model, and provides well constrained model parameters (see Table \ref{specpar}). We find the unabsorbed source flux to be $(5.88\pm0.02)\times10^{-11}$\,erg\,s$^{-1}$\,cm$^{-2}$ in the $0.5-6.5$\,keV range, and a temperature $kT=0.38\pm0.02$ keV, corresponding to a hot spot surface area of $131\pm27$\,km$^2$ (assuming a source distance of 2\,kpc).

Using the STEMS fit, we obtained a hydrogen column density of $N_{\rm H}=(6.7\pm0.2)\times10^{21}$\,cm$^{-2}$, that is much lower than the one obtained with the BB$+$PL fit ($N_{\rm H}=9.1\times10^{21}$\,cm$^{-2}$). The weighted average value of $N_{\rm H}$ towards the direction\footnote{As obtained with the HEASARC nH tool: http://heasarc.gsfc.nasa.gov/cgi-bin/Tools/w3nh/w3nh.pl} of \sgr is $6.2\times10^{21}$\,cm$^{-2}$ and $5.2\times10^{21}$\,cm$^{-2}$ using the \citet{dlmap} and the \citet{labmap} surveys, respectively. The $N_H$ from the STEMS fit agrees well with the Galactic value, however, the BB$+$PL model fit overestimates the absorption.

\subsection{The Time-integrated Burst spectrum}

\begin{figure}[h]
    \includegraphics[scale=0.8]{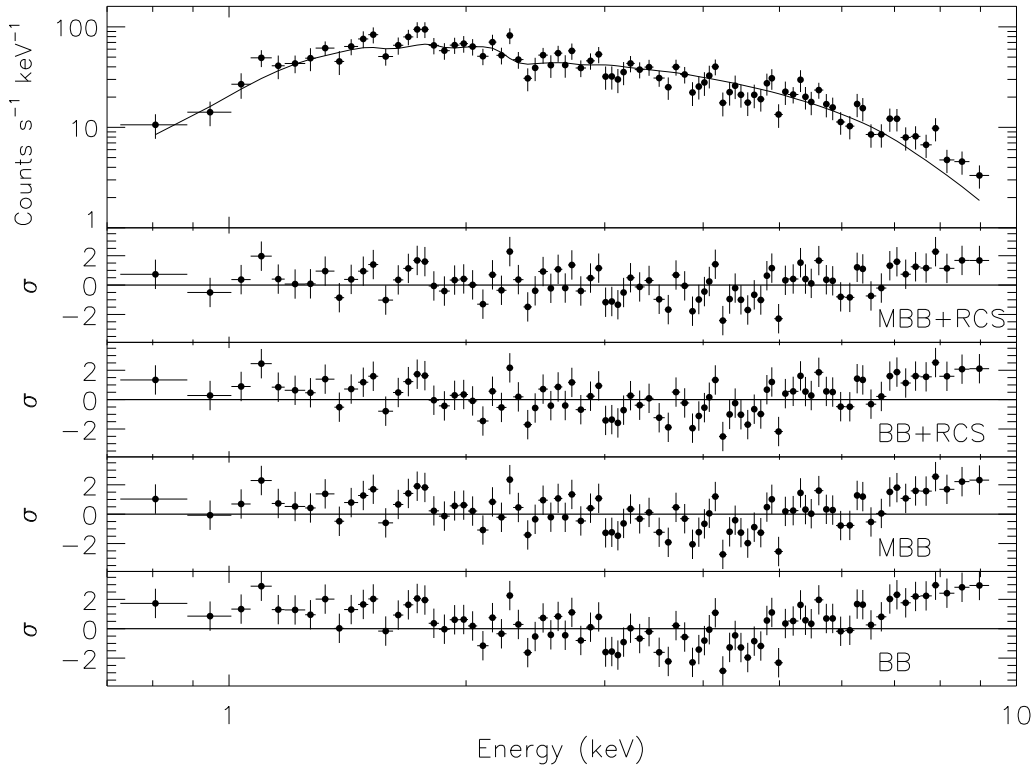}
\caption{The time-integrated stacked spectrum of 129 dim \sgr bursts overploted with the best fit MBB+RCS model and the residuals for MBB+RCS, BB+RCS, MBB, and BB models (from the top to bottom).  \label{specburstall}}
\end{figure}

To remove the contribution of the persistent emission from the burst emission, we used the persistent spectrum as the background for the burst spectrum. We fit the burst spectrum with several continuum models: blackbody, modified blackbody, blackbody with RCS, and MBB$+$RCS. In all our fits, we fixed $N_{\rm H}$ at $6.7\times10^{21}$\,cm$^{-2}$, the value obtained from the persistent spectrum fit with the STEMS model. Previous studies revealed that the magnetospheric parameters do not vary significantly over timescales of months \citep{guver2007,guver2008,gogus2011}. We also assumed that the RCS by the non-relativistic warm plasma in the magnetosphere of \sgr is not varying over the course of the \textit{XMM-Newton} observation and adopted the magnetospheric electron velocity  ($\beta$) and the optical depth ($\tau$) from the persistent spectrum fit. We list all the fit results in Table \ref{specpar}. Figure \ref{specburstall} exhibits the burst spectrum with the MBB$+$RCS fit (top panel) and the residuals for all models (next four panels). We find that a MBB$+$RCS gives the best fit among all four models. Note the apparent deviation above 7\,keV in all spectral fits; this may be due to an additional component in the spectra of SGR bursts as shown by \citet{lin2012}. We calculated the average unabsorbed flux of the bursts in the $0.5-10$\,keV to be $(1.80\pm0.05)\times10^{-9}$\,erg\,s$^{-1}$\,cm$^{-2}$. The total energy released by all 129 very dim bursts is $\sim$7.5$\times10^{36}$\,erg.

\subsection{The Time-resolved Burst spectrum}

\begin{figure}[h]
    \includegraphics{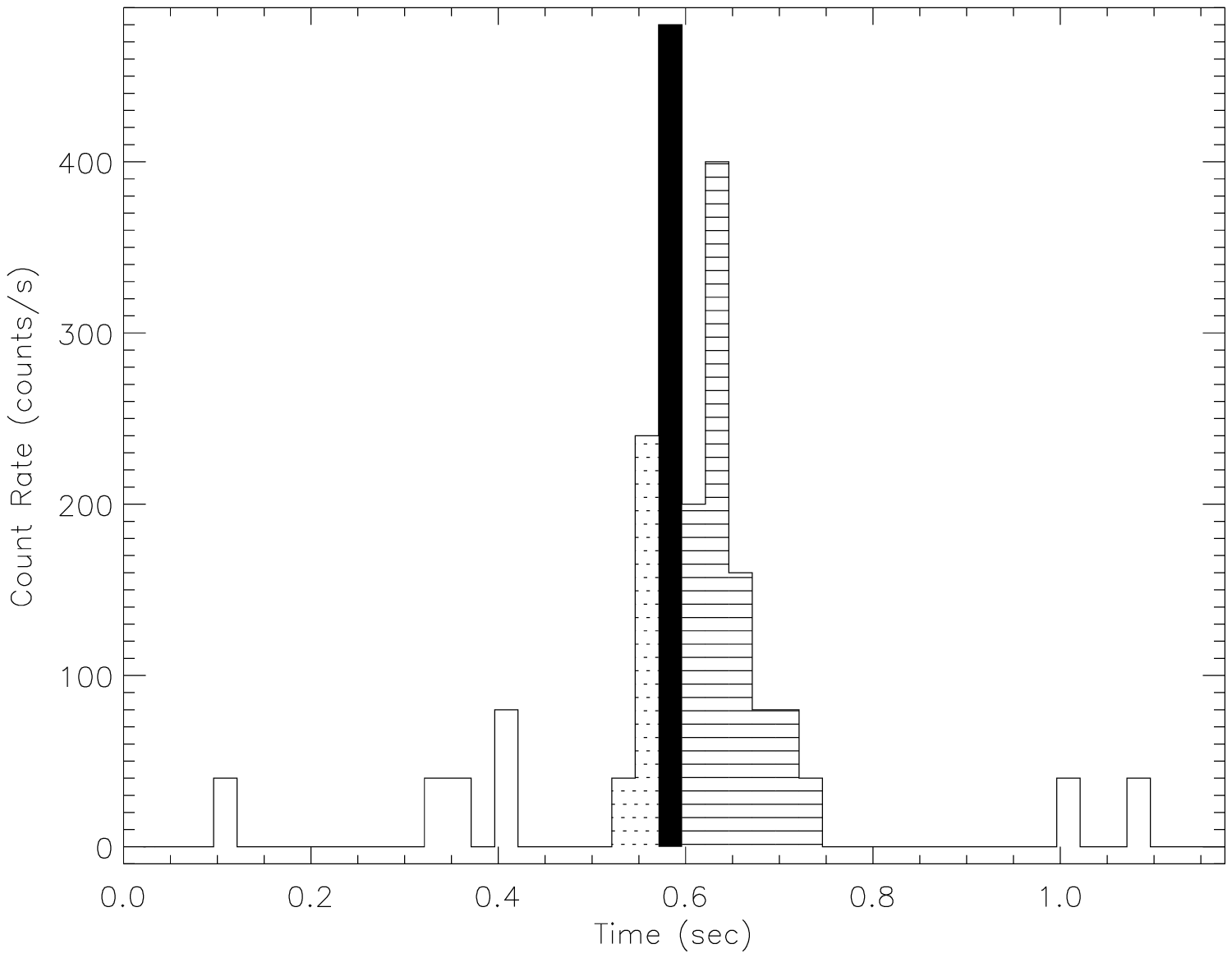}
\caption{An example of a bright burst lightcurve from \sgr with 25\,ms time resolution. The black bar is the peak bin. The rise and decay parts are filled with dots and horizontal histograms, respectively. The reference time of the x-axis is 2008 August 23 12:19:16.861 UTC. \label{trlc}}
\end{figure}

We studied the time evolution of the spectral properties of the brightest bursts in our set of 129 events, using their stacked time-resolved spectra. We selected all the bursts with peak count rate more than 300 counts\,s$^{-1}$ on a 25\,ms binsize. Note the fact that we reduced the time resolution to accommodate SGR bursts with finer temporal structures. This selection resulted in a set of 47 bursts. Then, we separated each selected burst into three parts: rise, peak, and decay. The peak bin is defined as the 25\,ms time bin with the highest count rate during the burst; the rise part is the time interval from the onset of the burst to the start of the peak bin; and the decay part starts from the end of the peak bin ending when the emission returns back to the background level. Figure \ref{trlc} shows an example of the three parts of a relatively bright burst. We generated stacked spectra for the rise, peak and decay parts and fit these spectra with the MBB$+$RCS model.

\begin{figure}[h]
    \includegraphics{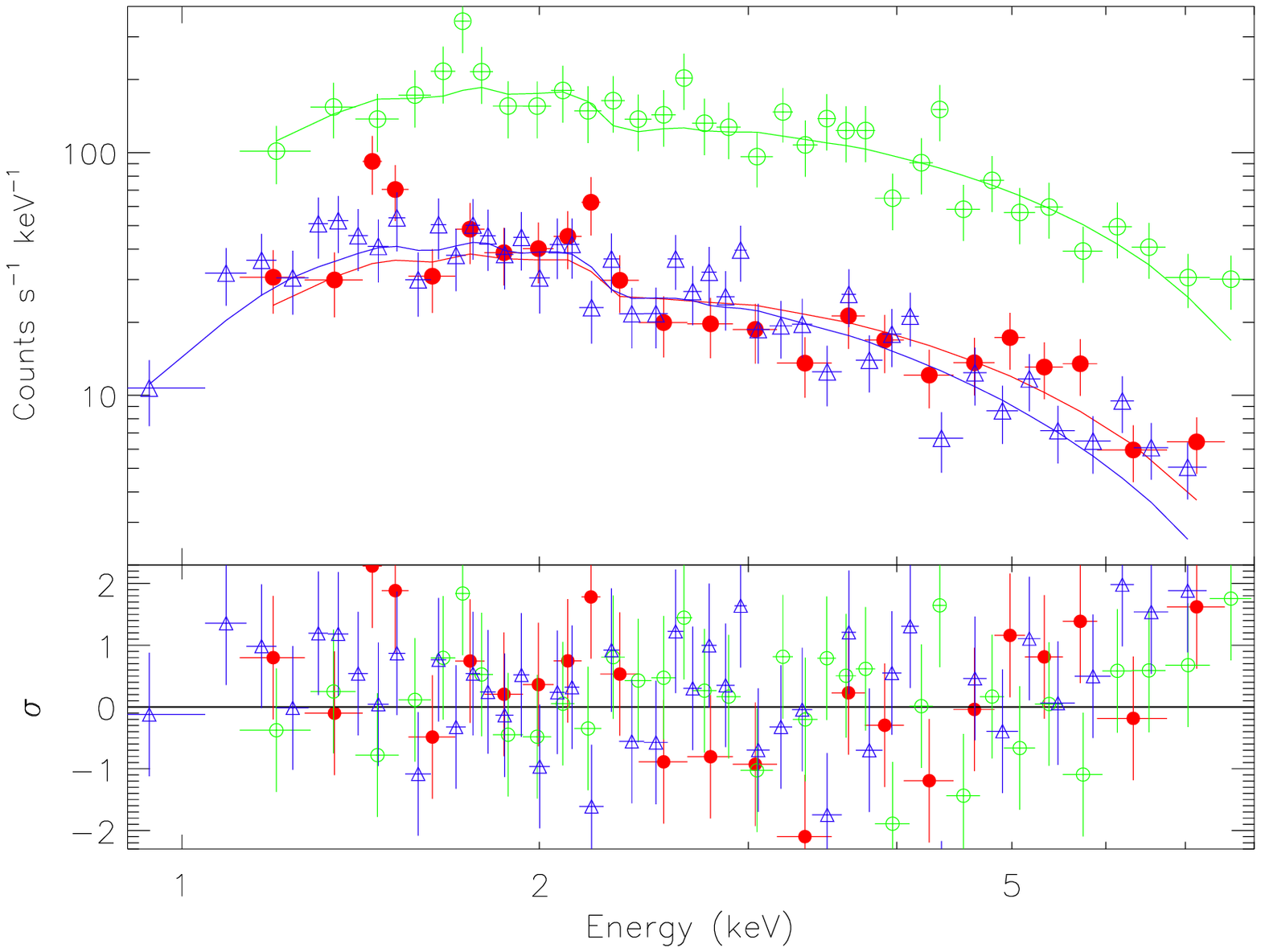}
\caption{The count spectra and their best fit with the MBB$+$RCS model curves for the rise (dots), peak (circles), and decay (triangles) parts of the stacked bursts. \label{trspec}} 
\end{figure}

\begin{figure}[h]
    \includegraphics[scale=0.8]{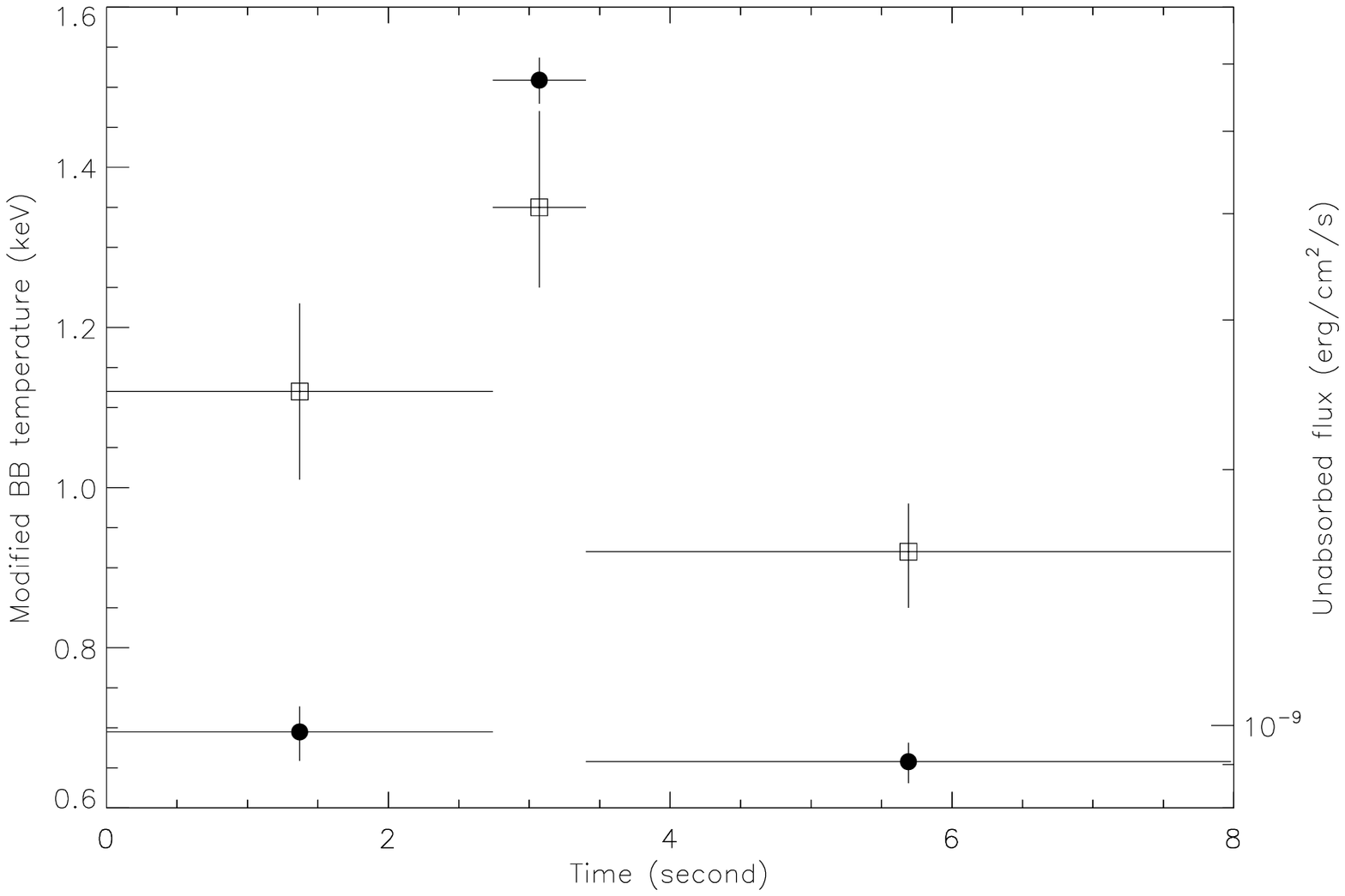}
\caption{Time evolution of the temperature of the modified blackbody (\textit{squares}) and the unabsorbed flux (\textit{filled circles}) in the $0.5-10$\,keV range for the rise, peak and decay parts of the 47 brightest bursts of \sgrnos. The error bars in the time axis represent the total exposure time of each accumulated part.  \label{ktevo}}
\end{figure}

Figure \ref{trspec} exhibits the spectra of the rise, peak and decay parts of the bursts, and their best fit model curves. We list all spectral fit results in Table \ref{specpar}. We find that the burst bubble temperature is the highest at the peak among the three parts. However, the temperatures of the rise and peak parts differ only by $\sim2\sigma$. Therefore, we fit these two parts simultaneously, forcing them to yield a single (best fit) temperature of $1.25\pm0.08$\,keV. We notice that the temperature of the modified blackbody is lower ($>3\sigma$) during the decay phase of the bursts. The average unabsorbed burst flux ($0.5-10$\,keV) is $(9.8\pm0.7)\times10^{-10}$\,erg\,s$^{-1}$\,cm$^{-2}$, $(5.7\pm0.4)\times10^{-9}$\,erg\,s$^{-1}$\,cm$^{-2}$, and $(9.1\pm0.5)\times10^{-10}$\,erg\,s$^{-1}$\,cm$^{-2}$ for the rise, peak and decay parts, respectively. We present a coarse time evolution of the burst temperature and flux in Figure \ref{ktevo}. 

\begin{deluxetable}{ccccccccc}
\tabletypesize{\scriptsize}
\setlength{\tabcolsep}{0.025in} 
\tablecaption{Spectral fit results of the persistent and burst emission from \sgrnos. \label{specpar}}
\tablewidth{0pt}
\tablehead{
\colhead{Spectrum} & \colhead{Model} & \colhead{$N_H$} & \colhead{$B$} & \colhead{$kT$\tablenotemark{a}} & \colhead{index} & \colhead{$\beta$} & \colhead{$\tau$} & \colhead{$\chi^2_{\nu}$/dof} \\
 & & ($10^{22}$\,cm$^{-2}$) & ($10^{14}$\,G) & (keV) & & & & 
}
\startdata
Persistent & BB+PL & $0.91\pm0.01$ & - & $0.70\pm0.01$ & $2.79\pm0.04$ & - & - & $0.7657/117$ \\
Persistent & STEMS & $0.67\pm0.02$ & $2.21\pm0.07$ & $0.38\pm0.02$ & - & $0.37\pm0.01$ & $5.0\pm0.2$ & $0.7615/116$ \\
\hline
Burst all & BB & $0.67$ & - & $1.08\pm0.02$ & - & - & - & $2.1054/84$ \\
Burst all & MBB & $0.67$ & - & $1.35\pm0.04$ & - & - & - & $1.4944/84$ \\
Burst all & BB+RCS & $0.67$ & - & $0.93\pm0.03$ & - & 0.37 & 5.0 & $1.4415/84$\\
Burst all & MBB+RCS & $0.67$ & - & $1.16\pm0.04$ & - & 0.37 & 5.0 & $1.174/84$\\
Burst rise & MBB+RCS & $0.67$ & - & $1.12\pm0.11$ & - & 0.37 & 5.0 & $1.019/20$\\
Burst peak & MBB+RCS & $0.67$ & - & $1.35^{+0.12}_{-0.10}$ & - & 0.37 & 5.0 & $0.9632/25$\\
Burst decay & MBB+RCS & $0.67$ & - & $0.92^{+0.06}_{-0.07}$ & - & 0.37 & 5.0 & $1.169/36$\\
Burst rise+peak & MBB+RCS & $0.67$ & - & $1.25^{+0.09}_{-0.07}$ & - & 0.37 & 5.0 & $1.015/46$\\
\enddata
\tablenotetext{a}{For all burst fits this parameter is the $kT_b$ of the modified blackbody model}
\end{deluxetable}

\section{Discussion \label{sec:disc}}

We investigated the persistent X-ray and very dim burst emission properties of \sgr during the day after the onset of its 2008 outburst, using the deepest XMM-Newton observations of the source that included a large number of bursts. We analyzed the persistent X-ray emission with the STEMS magnetar emission model and found a magnetar average surface temperature of $kT=0.38$ keV, that is much lower than the temperature of $\sim1$\,keV, obtained with the phenomenological BB$+$PL model fit. The persistent emission of almost all other sources modeled with BB$+$PL yields a blackbody temperature of $\sim$0.5\,keV. Note that this value is much lower than that of \sgr fit with BB$+$PL, but closer to or slightly higher than that estimated with the STEMS model fit. We also found that for \sgrnos, the BB$+$PL model gives a larger $N_{\rm H}\sim10^{22}$\,cm$^{-2}$ than the Galactic value (5$-$6$\times$10$^{22}$\,cm$^{-2}$) in the direction towards \sgrnos, while the STEMS $N_{\rm H} = 6.7\times10^{21}$ cm$^{-2}$ agrees well with the latter value. 

\citet{enoto2010} investigated the \textit{Suzaku} observations of \sgr on 2008 August $26-27$ in the energy range of $1-70$\,keV. They used a thermal emission with a resonant cyclotron scattering model \citep[see ][]{lyutikov2006} to describe the persistent soft X-ray spectrum plus a simple power law for the hard X-ray component. This model did not provide a good fit (reduced $\chi^2=1.73$). They reported a value for $N_{\rm H}=6.2\times10^{21}$\,cm$^{-2}$, a surface temperature of $kT=0.39$\,keV, and magnetospheric scattering parameters, $\tau=5.1$, and $\beta=0.3$, which are quite similar to our results. However, the uniqueness of the STEMS model is the fact that it takes into account the presence of the strong surface magnetic field and its possible effect on the emergent spectrum via atmospheric processes. Using STEMS we obtain, for the first time, the surface magnetic field strength of \sgr via X-ray spectroscopy, as $B=2.2\times$10$^{14}$ G, a value which agrees well with the inferred equatorial dipole field strength of 2$\times$10$^{14}$ G \citep{rea2009,gogus2010}.

\begin{figure}[h]
    \includegraphics[scale=0.8]{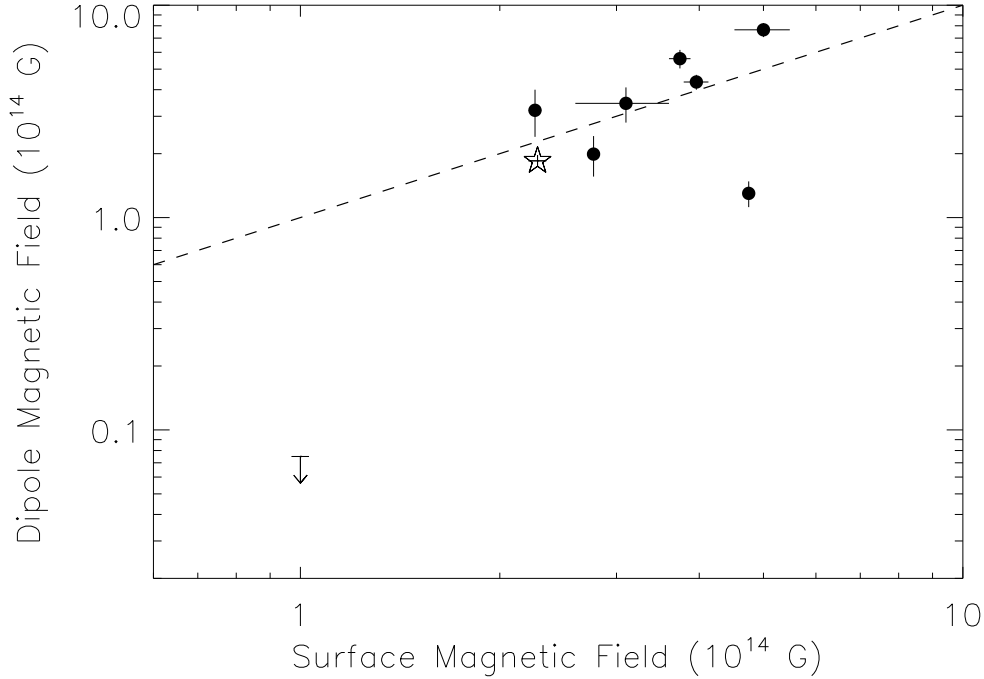}
\caption{The comparison of the inferred dipole magnetic fields with the surface magnetic field obtained with X-ray spectroscopy for nine magnetars. The dashed line marks the line on which the two axes are equal. \sgr is shown as a five-point star. All other values in this figure are taken from \citet{guver2011}. SGR\,J$0418+5729$ has an upper limit for the inferred dipole magnetic field and a measurement with STEMS spectroscopy.  \label{Bfield}}
\end{figure}

The STEMS model has already been used to study the persistent emission spectra of four AXPs, XTE\,J$1810-197$, 4U\,0142$+$61, 1E\,$1048.1-5937$ and 1RXS\,J$170849.0-400910$ \citep{guver2007,guver2008,ozel2008}, and four SGRs, SGR\,$1900+14$, SGR\,$0418+5729$, SGR\,$0526+66$ and SGR\,J$1550-5418$ \citep{gogus2011,guver2011,guver2012,ng2011}. In most cases, including \sgrnos, the surface magnetic field obtained from X-ray spectroscopy is consistent with the dipole magnetic field inferred from the spin period and the spin down rate, as shown in Figure \ref{Bfield}. Because magnetars emit energetic bursts, their surface magnetic field strengths are expected to be in the $10^{14}-10^{15}$\,G domain. Some sources, however, may have lower inferred dipole magnetic field strengths \citep{rea2010}, although higher order magnetic structures (multipolar fields) may be present \citep{guver2011}. The similarity between the surface magnetic field strengths of \sgr as obtained via X-ray spectroscopy and as inferred from the spin properties suggests that its surface magnetic field topology is close to a dipole.

We also studied here, for the first time, the dimmest bursts and their physical link to the persistent X-ray emission as both kinds of emission are redistributed by resonant cyclotron scattering process in the magnetosphere. The burst spectrum can be described with a warm bubble trapped by the magnetosphere, emitting thermal radiation modified by a strong magnetic field. The bolometric temperature $T_b$ is introduced by \citet{lyubarsky2002} with the total radiation flux as $F=\frac{1}{2}\sigma T_b^4$. The emission area of the burst can be approximately calculated as 
\begin{eqnarray}
A = \frac{4 \pi D^2 f}{\frac{1}{2}\sigma T_b^4} 
\end{eqnarray}
where $A$ is the emission area, $D$ is the distance to \sgrnos, $f$ is the unabsorbed flux, $\sigma$ is the 
Stefan-Boltzmann constant, and $T_b$ is the bolometric temperature. Assuming a distance to the source of 
$\sim2$\,kpc, we estimate the average emission area of the burst as $93\pm10$\,km$^2$, which is $\sim7.4\%$ 
of the magnetar surface ($R_{NS} = 10$\,km).

 \citet{enoto2010} and \citet{nakagawa2011} studied the spectra of the persistent emission and the dim short bursts (average flux $\sim 10^{-9}$\,erg\,s$^{-1}$\,cm$^{-2}$) from \sgr with \textit{Suzaku}. They fit both spectra with two blackbody functions and a power law model. Comparing the parameters of these fits,  \citet{nakagawa2011} concluded that the persistent emission and the dim bursts have the same emission mechanisms. They further suggested that the persistent emission of magnetars may be composed of numerous microbursts. We find that both kinds of emission are forms of modified thermal emission and redistributed by the resonant cyclotron scattering in the magnetosphere. However, the spectral shapes of the thermally dominated portion ($\lesssim$4 keV) of the STEMS and MBB+RCS models are remarkably different and they yield significantly different temperatures. We, therefore, argue against the idea that the persistent and burst emissions have a common origin. The main difference between the two is that the scattering is the dominant process in the burst bubble, while for the persistent emission, one needs to consider not only the scattering but also the emission and absorption processes as photons travel through the magnetar atmosphere. We investigated the dependence of burst peak times on the rotational phases of the source and we found that bursts occur quite uniformly over the entire spin phase, showing no significant concentration during any phases. This result supports our argument that short bursts and persistent emission do not have a common origin. 

Finally we investigated the stacked time-resolved spectra of the brighter bursts from \sgrnos. We found 
that their temperature traces their flux behavior and has a lower value in the decaying part compared to 
the rise and peak parts. We note here that these results do not contradict the earlier results by 
\citet{lin2011}, who find the burst hardness to evolve from hard-soft-hard with flux over a very broad 
burst flux range, since the bursts studied here correspond to a very narrow flux range in a softer energy 
band. From the Equation 1 above, we calculate the emission area for the rise, peak and decay parts as 
$58\pm19$\,km$^2$, $160\pm41$\,km$^2$, and $118\pm29$\,km$^2$, respectively. The emission area during the peak of the burst is similar to the area of the persistent emission, within $1\sigma$. Although more coarse, the behavior of cooling during the decaying tail is similar to what has been seen in the tails of energetic bursts from SGR\,$1900+14$ \citep{lenters2003} and SGR \,$1806$-$20$ 
\citep{gogus2011}, implying that cooling takes place at all energy scales of SGR events. 

\acknowledgments 

We thank Yuri Lyubarsky for stimulating discussions. L.L. is funded through the Post-Doctoral Research Fellowship of the Turkish Academy of Sciences (T\"UBA).

\end{document}